\documentclass[]{spie}  %>>> use for US letter paper
\usepackage{amsmath,amsfonts,amssymb}
\usepackage{graphicx}
\usepackage[colorlinks=true, allcolors=blue]{hyperref}
\usepackage{float}
\usepackage{listings}

\title{Tunable Kernel-Nulling interferometry for direct exoplanet detection}

\author[a,*]{Vincent Foriel}
\author[a]{Frantz Martinache}
\author[a]{David Mary}
\affil[a]{Université Côte d’Azur, Observatoire de la Côte d’Azur Nice, CNRS, Laboratoire Lagrange, Nice, France}

\authorinfo{Further author information:\\Vincent Foriel - E-mail: vincent.foriel@oca.eu\\  Frantz Martinache - E-mail: frantz.martinache@oca.eu\\ David Mary - E-mail: david.mary@oca.eu}

\begin{document}
\maketitle

\begin{abstract}

    Nulling interferometry\cite{Bracewell} is a promising technique for direct detection of exoplanets. However, the performance of current devices is limited by different perturbations sources \cite{Lay et al. 2004} and especially by its sensitivity to any phase aberrations. The work presented here attempts to overcome those limitations by using a four-telescopes nulling interferometer architecture, called Kernel-Nuller\cite{Martinache et al. 2018}, which includes a recombiner that positions the four signals in phase quadrature. This architecture is based on an integrated optical component containing 14 electronically controlled phase shifters, used to correct optical path differences that would be induced by manufacturing defects. The first part of the study consists in the development of an algorithm providing the delays to be injected into the component to optimize the performance of that device. The next step of this study deals with the analysis of the intensity distributions produced at the output of the Kernel-Nuller\cite{Martinache et al. 2018, Cvetojevic et al. 2022} through a series of observations. Then we apply statistical tests and data treatment techniques to detect the signature of an exoplanets.

\end{abstract}


\section{INTRODUCTION}
\label{sec:intro} % \label{} allows reference to this section

The direct detection of exoplanets is very challenging due to the high contrast between the stellar and planetary signals. Nulling interferometry offers a promising solution to this problem. This technique involves pointing multiple telescopes toward the target star to achieve destructive interference, effectively canceling out the star's light. Due to the angular separation between the star and its companion, the light from the latter reaches each telescopes with a different phase, preventing complete destructive interference and allowing for potential constructive interference of the planetary light. However, such a system is highly sensitive to phase aberations that limit their performances. Kernel-nulling is an attempt to partially overcome this issue by employing three or more\cite{N telescope kernel} telescopes to generate symetric pairs of outputs that can be subtracted to eliminate low-order phase aberrations\cite{Martinache et al. 2018}. This study aims to enhance the performance of such a system by introducting phase aberation control techniques and statistical analysis to improve the detection of exoplanets.

\section{Architecture}

\begin{figure} [H]
    \begin{center}
    \begin{tabular}{c}
    \includegraphics[height=6cm]{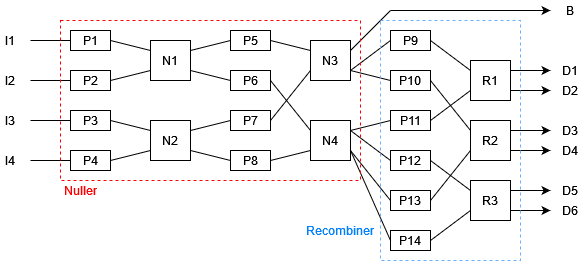}
    \end{tabular}
    \end{center}
    \caption[kernel_nuller] 
    { \label{fig:kernel_nuller} 
    Schematic of the tunable Kernel-Nuller architecture. P1-14 are the thermo-optic phase shifters. N1-4 are the four $2 \times 2$ Nuller MMIs, forming a $4 \times 4$ arrangement. R1-3 are the three recombiners that place the signals in phase quadrature. I1-4 are the input signals. B is the bright output and D1-6 are the dark outputs. Cameras are placed at B and D1-6.}
\end{figure}

\begin{figure} [H]
    \begin{center}
    \begin{tabular}{c}
    \includegraphics[height=3cm]{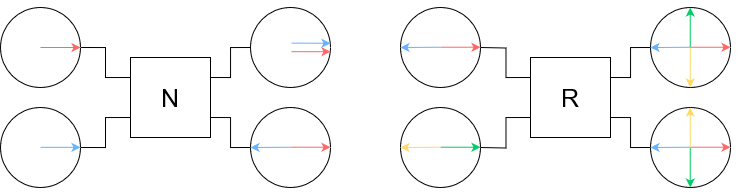}
    \end{tabular}
    \end{center}
    \caption[nuller_and_recombiner] 
    { \label{fig:nuller_and_recombiner} 
    Schematic of the action of nuller (left) and recombiner (right) MMIs. The signals (electric fields) are represented by arrows in a complex polar plot showing the phase and amplitude of the signal. As we are insensitive to absolute phase, one of the signals (here the red one) is used as a reference.}
\end{figure}

This study employs a 4-telescope Kernel-Nuller architecture, using integrated optical components. The interference is achieved through a series of $2 \times 2$ multimode interferometers\cite{MMI} (MMI). There are two types of MMIs in this system. The first type, the nullers, denoted $N_n$, perform constructive interference at the first output and destructive interference at the second output by placing the two electric fields in phase opposition (Equation \ref{N} \& Figure \ref{fig:nuller_and_recombiner}). The second type, referred as "recombiners", denoted $R_n$, are placed after the nuller one and allow to put the four signals in phase quadrature (Equation \ref{R} \& Figure \ref{fig:nuller_and_recombiner}). These MMIs are separated by thermo-optic phase shifters, denoted $P_n$, that allow us to introduce phase shifts at various points in the system. The overall architecture is shown in Figure \ref{fig:kernel_nuller}.

The system has seven outputs: one bright output containing all four signals in phase synchronization, and six "dark" outputs containing combinations of signals in phase quadrature (Figure \ref{fig:phases}). Interestingly, these dark outputs occur in symmetric pairs, meaning that a disturbance in one output will similarly affect the other output in the pair. This allows to cancel low-order phase aberations by subtracting the two dark outputs in each pair, creating a new observable called "kernels"\cite{Martinache et al. 2018}.

Formally, the Nuller and Rcombiner MMIs can be expressed using the following matrices:

\begin{equation}\label{N}
    N_n =
    \begin{pmatrix}
        1 & 1 \\
        1 & -1
    \end{pmatrix}
\end{equation}

\begin{equation}\label{R}
    R_n =
    \begin{pmatrix}
        e^{i\frac{\pi}{4}} & e^{-i\frac{\pi}{4}} \\
        e^{-i\frac{\pi}{4}} & e^{i\frac{\pi}{4}} \\
    \end{pmatrix}
\end{equation}

\begin{equation}\label{P}
    P_n = e^{i\theta_n}
\end{equation}

\section{Calibration}

\begin{figure} [H]
    \begin{center}
    \begin{tabular}{c}
    \includegraphics[height=2.5cm]{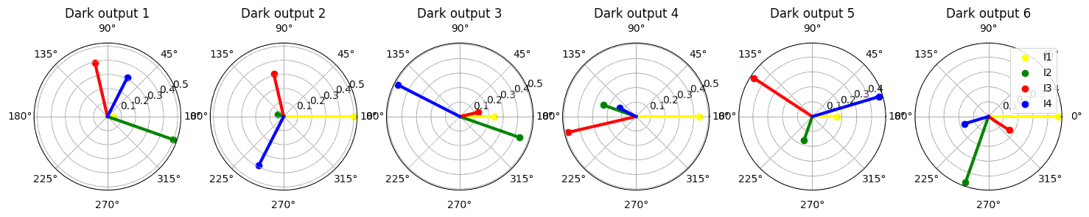}\\
    \includegraphics[height=2.5cm]{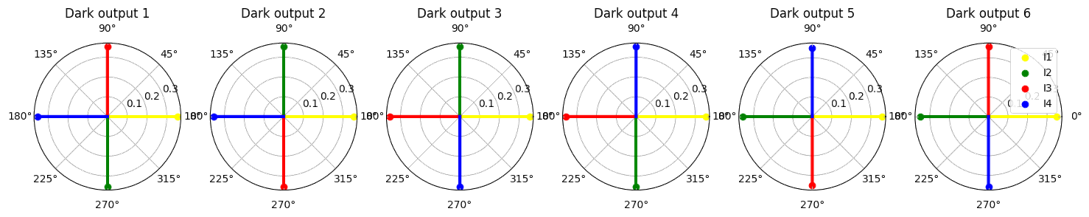}
    \end{tabular}
    \end{center}
    \caption[phases] 
    { \label{fig:phases} 
    Polar representation of the complex amplitudes obtained at the dark outputs. Each output is a combination of signals from the 4 telescopes, ideally separated by 90°. The first row is what we get without calibrating the system. We observe that phase perturbations of the order of $\lambda / 10$ cause significant performance degradation. However, as shown on the bottom row, the calibration algorithm effectively corrects these perturbations, restoring the signals to the desired phase quadrature. These results were observed over a large number of simulations with different aberations.}
\end{figure}

Due to unavoidable manufacturing defects, the system will produce some phase aberrations that we aim to cancel using the active thermo-optic phase shifters. There are fourteen phase shifters in total, and we need to find the optimal values to place the signals in phase quadrature (see Figure \ref{fig:phases}) and optimize system performance. To achieve this, we use a deterministic genetic algorithm to explore the parameter space and find these optimal values. This algorithm consists in interatively changing (mutation operation) one of the phase shifters in a specific order and evaluate the impact on the system performance. If the performance improves, the change is kept, otherwise it is discarded. This process is repeated for each phase shifter in a loop, reducing the modification step size each time. The algorithm stops when the modification step size is below a certain threshold, which will be determined by the physical limitations such as the photon noise or the precision with which we can adjust the phase shifters.

To evaluate the performance, we consider a single star without phase perturbation at the input. In this case, all the light flux should be directed to the bright output, and all kernels should be perfectly null. We use two metrics ($M_B$ and $M_K$ below) to evaluate system performance, based on the electric fields at the bright ($B$) and dark ($D_n$) outputs:

\begin{equation}
    M_B = |B|^2
\end{equation}

\begin{equation}
    M_K = \Big||D_1|^2 - |D_2|^2\Big|+\Big||D_3|^2 - |D_4|^2\Big|+\Big||D_5|^2 - |D_6|^2\Big|
\end{equation}

This algorithm will then try to maximize $M_B$ and minimize $M_K$. To do so, each phase shifter is associated with one of these performance metric. Phase shifters $P_1$ to $P_5$ and $P_7$ (see Figure \ref{fig:kernel_nuller}) are associated with the $M_B$ metric as they contribute to redirecting the flux to the bright output. The remaining phase shifters are associated with the $M_K$ metric. For each output we take the square modulus as we only get the intensity of the resulting electric fields.

\section{Performances and limitations}

\begin{figure} [H]
    \begin{center}
    \begin{tabular}{c}
    \includegraphics[height=5cm]{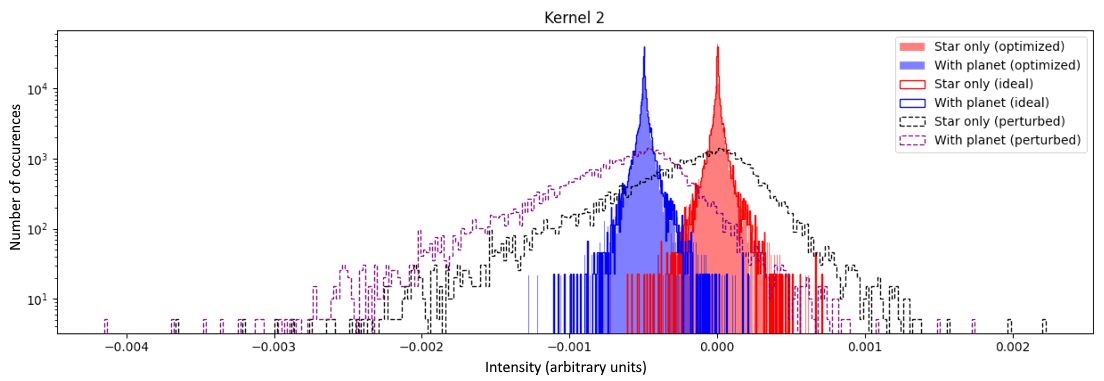}
    \end{tabular}
    \end{center}
    \caption[distrib] 
    { \label{fig:distrib} 
    Intensity distributions obtained on the second kernel for both hypotheses (with and without a planet). We see that distributions obtained after the calibration process, represented by the colored filled parts, closely match the distributions expected with an ideal component, represented by the continuous lines. To clearly see the distribution shift due to the presence of a planet, the contrast has been set to $10^{-3}$. The RMS phase perturbation is set to $\lambda / 100$, so $16.5$nm, which is optimistic for ground telescopes (~100nm), but pessimistic for a space application (~1nm).}
\end{figure}

The calibration corrects phase errors caused by manufacturing defects but does not address phase errors occurring upstream, such as those induced by the atmosphere. For each observation, we obtain an intensity value for each kernel that is a realisation of a random variable  distributed around this value, with a dispersion that is directly dependent on the phase perturbations (see Figure \ref{fig:distrib}). If a planet is present, the intensity distribution will shift compared to the expected distribution without a planet. 

\subsection*{Performance evaluation}

\begin{figure} [H]
    \begin{center}
    \begin{tabular}{c}
    \includegraphics[height=5cm]{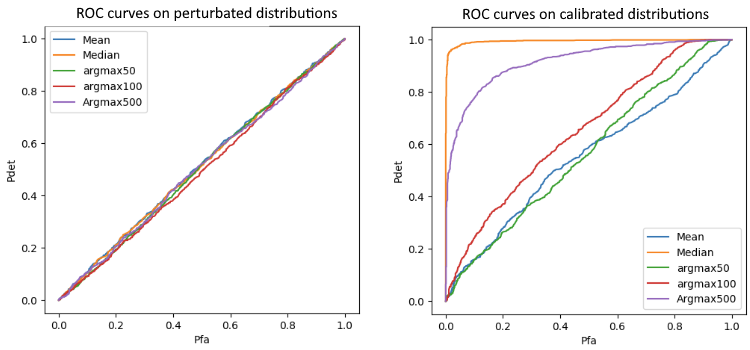}
    \end{tabular}
    \end{center}
    \caption[ROC] 
    { \label{fig:ROC} 
    These plots shows the detection performance (Pdet) as a function of the probability of false alarm (Pfa) for different methods. The methods used here consist of reducing the distribution to a single value using the median, mean, and argmax (with different number of bins) operations and comparing this value to a threshold. The higher the threshold, the more the null hypothesis is favored. The lower the threshold, the more false alarms we encounter. So, by varying the threshold, we obtain for each considered test statistic different trade-offs Pdet vs Pfa, the set of which draws a curve called ROC curve. These results were obtained using 1000 different distributions containing 1000 intensities each. We clearly see that, in these conditions, the calibration process leads to more powerful statistical tests. Moreover, among the considered test statistics, the median appeared to be the most powerful.}
\end{figure}

As the signal to noise ratio is very low, the shift is not trivial to observe. We then use statistical tests to evaluate the presence of such a shift (and then the presence of a planet). So far, only simple tests have been conducted (see Figure \ref{fig:ROC}), reducing the distribution to a single intensity value. The considered test statistics are the mean, the median, and different argmax (position of the highest bin, "argmax100" refer to a 100 bin sampling) of the obtained distribution and comparing this intensity to a threshold value. If this latter value is too low, we will often detect a shift that is actually due to the phase perturbations, and then the probability of false alarm will be high. If it is too high, we will not detect the planet, so the probability of detection will be low. We can estimate the power of the test by comparing the probability of detection to the probability of false alarm (Figure \ref{fig:ROC}).

For a series of observations performed in the same conditions, looking at the median of the distribution is theoretically sufficient to detect planets with a contrast of $10^{-5}$ in the presence of phase perturbations of the order of $\lambda / 100$ in 90\% of cases with a 1\% probability of false alarm. These results tends to confirm the accuracy already reached in the literature \cite{Cvetojevic et al. 2022}, but there is still a chance to improve them by studying more advanced statistical tests.

\section{Position caracterization}

\begin{figure} [H]
    \begin{center}
    \begin{tabular}{c}
    \includegraphics[height=5cm]{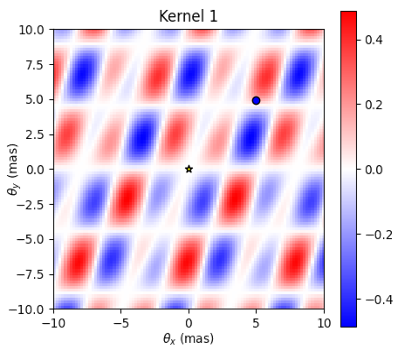}
    \end{tabular}
    \end{center}
    \caption[transmission_map] 
    { \label{fig:transmission_map} 
    Depending on the telescope positions, the signals of each object with an certain angle from the line of sight will arrive with a phase differences that either constructively or destructively interfere. This phenomenon leads to a transmission map, different for each kernel. As kernels are built using the difference in intensity of two dark outputs, this transmission map can go negative. If a planet is positioned in a red zone, the obtained distribution (Figure \ref{fig:distrib}) will shift towards positive values, and if it is in a blue zone, it will shift towards negative values.}
\end{figure}

\begin{figure} [H]
    \begin{center}
    \begin{tabular}{c}
    \includegraphics[height=5cm]{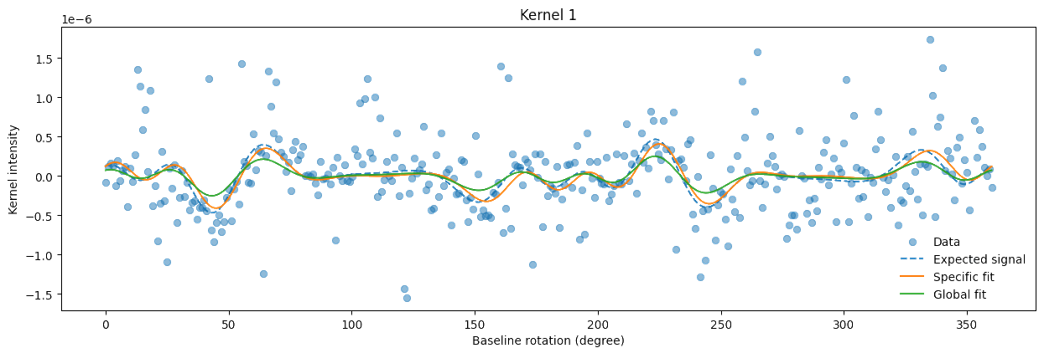}
    \end{tabular}
    \end{center}
    \caption[fit] 
    {\label{fig:fit} 
    To determine the position of the planet, we fit the signal modulation function to the obtained data using the non linear fitting algorithme included in the \lstinline{scipy.optimize} python pakcage. This fitting process is conducted for each kernel individually, referred as the "specific fit". After performing the specific fits, we average the resulting positions to derive a new position estimate. We then plot the modulation for this averaged position, referred to as the "global fit". In practice, as we do not know the expected signal, the correlation of the global fit and each specific fits gives us some information about the goodness of these fits. In this simulation example, we set a contrast of $10^{-5}$ and a RMS phase perturbation of $\lambda / 100$ and we only consider a simple baseline rotation, resulting in a rotation of the transmission map. In practice, due to the earth shape, the projection of the baseline during the night vary in a more complex way, leading to a more complex modulation function that can be difficult to fit properly.}
\end{figure}

Beyond detecting the presence of an exoplanet, this system can caracterize its position by performing a series of observations at various hour angles. Using the transmission maps of the different kernels (on-sky projection of the outputs, see Figure \ref{fig:transmission_map}) for each different hour angle \cite{Chingaipe}, we can determine the expected signal modulation for a planet at a given location. By fitting this function to the obtained data using a non-linear least square method, we can determine the planet's position (Figure \ref{fig:fit}).

The modulation function can be obtained by considering the value of a point in the transmission map over an entire revolution of this map around its center. As the transmission map is highly repetitive, a non-linear least square fitting algorithm - such as the one provided by scipy - can easily fall into local minimums. It then often lead to incorrect results unless it is initially given an approximate but close position and contrast to the expected true parameters. Unfortunately, we lack this information. One approach is to perform the fit by using each point in the position and contrast parameter grid as an initial parameter and identifying when the fit properly converge.

\begin{figure}[H]
    \begin{center}
    \begin{tabular}{c}
    \includegraphics[height=3.5cm]{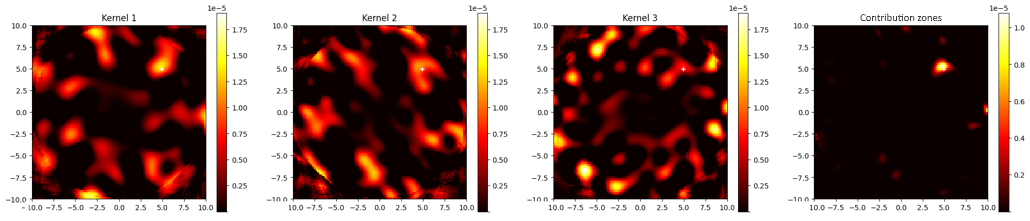}
    \end{tabular}
    \end{center}
    \caption[contribution_zone] 
    {\label{fig:contribution_zone} 
    Each of the first three plots represents the sum of the contribution maps at different times, weighted by the data obtained for that time. We are still in the particular case where the hour angle is approximated to a simple baseline rotation. The final plot represents the product of the first three plots which show here the approximate position of the planet we placed in the simulation (white cross). These results where also obtained for a contrast of $10^{-5}$ and a RMS phase perturbation of $\lambda / 100$. Also, this process can be considered for more powerfull statistical tests.}
\end{figure}

An alternative solution is to generate a sky map highlighting the regions that could potentially have contributed to the observed data. This method, already explored with classic nulling interferometers\cite{Image reconstruction}, involves weighting each rotated transmission map by the data obtained for that specific parallactic angle (detailed below), then summing all the weighted maps. It results in a map that highlight the planet's position (Figure \ref{fig:contribution_zone}).

The base idea was already explored as "image reconstruction" technic using classical nulling interferometry $^1$. However, the method here is based on Kernel-Nulls which have asymetric transmission maps. Thus, allow to better constrain the source of the signal by filtering the negative contributions.

Considering:
\begin{itemize}
\item $T_{n}$ represents the $n$-th kernel's transmission map normalized such as the max is 1.
\item $d_{n,\beta}$ denotes the data point obtained for kernel $n$ with baseline rotation $\beta$. Ex: $d_{1,\beta} = |D_{1,\beta}|^2 - |D_{2,\beta}|^2$ (see figure \ref{fig:kernel_nuller}).
\item $h$ is the hour angle.
\item $\theta$ is the angular separation.
\end{itemize}

We can define the raw contribution zones for each kernel at a given hour angle:

\begin{equation}
    \label{eq:raw_kernel}
    r_n(\theta, \alpha) = \sum_h T_{n,h}(\theta,\alpha) d_{n,h}
\end{equation}

As the kernel outputs are antisymetric, we can filter the negative contributions:
\begin{equation}
    \label{eq:heaviside}
    r'_n = \max(0, r_n)
\end{equation}

Finally, we can compute the product over all the kernels to get the final contribution zones:
\begin{equation}
    \label{eq:final}
    C(\theta, \alpha) = \prod_n r'_n(\theta, \alpha)
\end{equation}

\section{Discussion \& Prospects}

These simulations and statistical analysis shows highly promising results since we reach a detection contrast of $10^{-5}$ to $10^{-6}$ while the null depth goes up to $10^{-3}$ in the litterature \cite{Cvetojevic et al. 2022}. It also indicates the potential for even greater contrast by leveraging angular diversity.

Several factors could reduce these performances, such as amplitude aberrations, the chromatic nature of light, or the presence of multiple objects around the star. Adding these constraints to the simulation model and conducting an in-depth study on possible statistical tests will help to better define the actual limitations of such a system.

The photonic chip is already available and will be tested in the near future. We will then be able to compare the results obtained in the laboratory with those obtained in the simulations presented here.

\acknowledgments

Thanks to Romain Laugier for his wise advice, to Nick Cvetojevic for his introduction to the topic of photonics and to Margaux Abello for her presentation recommendations. This thesis is made possible by the PHOTONICS project of the PEPR ORIGINS and Thales Alenia Space.

% References
% \bibliography{report} % bibliography data in report.bib
% \bibliographystyle{spiebib} % makes bibtex use spiebib.bst

\end{document}